\documentclass[reprint,
amsmath,amssymb,11pt,reprint,aps,prl,showpacs,floatfix]{revtex4-2}

\usepackage{lineno}
\usepackage[english]{babel}
\usepackage{multirow}
\usepackage{xspace}
\usepackage{ulem}
\usepackage{soul}
\usepackage{graphicx}
\graphicspath{{figures//}} 
\usepackage{dcolumn}
\usepackage{bm}
\usepackage{color}
\usepackage{wrapfig}
\usepackage{hyperref}    
\usepackage{array}
\usepackage{booktabs}
\usepackage{multirow}
\usepackage{amssymb}

\usepackage{marvosym}
\hypersetup{
    unicode=false,          
    pdftoolbar=true,        
    pdfmenubar=true,        
    pdffitwindow=false,     
    pdfstartview={FitH},    
    pdftitle={My title},    
    pdfauthor={Author},     
    pdfsubject={Subject},   
    pdfcreator={Creator},   
    pdfproducer={Producer}, 
    pdfkeywords={keyword1} {key2} {key3}, 
    pdfnewwindow=true,      
    colorlinks=true,       
    linkcolor=blue,          
    citecolor=blue,        
    filecolor=blue,      
    urlcolor=blue,       
    plainpages=false        
}

\newcommand{\etal}{\textit{et al.}\xspace}

\newcommand{\TSDW}{$T_{\mathrm{{SDW}}}$\xspace}

\newcommand{\BKFA}{\mbox{Ba$_{1-x}$K$_x$Fe$_{2}$As$_2$}\xspace}

\newcommand{\LNO}{\mbox{La$_3$Ni$_2$O$_7$}\,\xspace}
\newcommand{\Alg}{$A_{\rm{1g}}$\,\xspace}
\newcommand{\Ag}{$A_{\rm{g}}$\xspace}

\newcommand{\Blg}{${B_{\rm{1g}}}$\xspace}
\newcommand{\BZg}{${B_{\rm{2g}}}$\xspace}

\newcommand{\wn}{\rm{cm}$^{-1}$\,}

\begin{document}

\title{\Large Anisotropic Electronic Correlations in the Spin Density Wave State of La$_3$Ni$_2$O$_7$}

\author{Ge He$^{1,2,*}$\textrm{\Letter}, 
Jun Shen$^{1}$\textrm{\Letter},
Shiyu Xie$^{3,*}$,
Haotian Zhang$^{1}$,
Mengwu Huo$^{4}$,
Jun Shu$^{5}$,
Deyuan Hu$^{4}$,
Xiaoxiang Zhou$^{6}$,
Yanmin Zhang$^{7}$,
Lei Qin$^{7}$,
Liangxin Qiao$^{3}$, 
Hengjie Liu$^{3}$, 
Chuansheng Hu$^{3}$, 
Xijie Dong$^{5}$,
Dengjing Wang$^{5}$,
Jun Liu$^{1}$,
Wei Hu$^{8}$,
Jie Yuan$^{8, 9}$,
Yajun Yan$^{6}$,
Zeming Qi$^{3}$,
Kui Jin$^{8, 9, 10}$, 
Zengyi Du$^{6}$\textrm{\Letter},
Meng Wang$^{4}$,
Donglai Feng$^{6}$\textrm{\Letter}}

\affiliation{
$^1$ School of Mechanical Engineering\mbox{,} Beijing Institute of Technology\mbox{,} Beijing 100081\mbox{,} China \\
$^2$ Beijing Key Laboratory of Quantum Matter State Control and Ultra-Precision Measurement Technology\mbox{,} Beijing Institute of Technology\mbox{,} Beijing 100081\mbox{,} China\\
$^3$ National Synchrotron Radiation Laboratory\mbox{,} University of Science and Technology of China\mbox{,} Hefei\mbox{,} Anhui 230029\mbox{,} China\\
$^4$ Center for Neutron Science and Technology\mbox{,} Guangdong Provincial Key Laboratory of Magnetoelectric Physics and Devices\mbox{,} School of Physics\mbox{,} Sun Yat-Sen University\mbox{,} Guangzhou 510275\mbox{,} China\\
$^5$ Department of Applied Physics\mbox{,} Wuhan University of Science and Technology\mbox{,} Wuhan 430081\mbox{,} China\\
$^6$ Hefei National Laboratory\mbox{,} and New Cornerstone Science Laboratory\mbox{,} Hefei, Anhui 230088\mbox{,} China\\ 
$^7$ Beijing Key Laboratory for Sensor, Beijing Information Science and Technology University, Beijing 100192, China\\
$^8$ Beijing National Laboratory for Condensed Matter Physics\mbox{,} Institute of Physics\mbox{,} Chinese Academy of Sciences\mbox{,} Beijing 100190\mbox{,} China\\
$^9$ School of Physical Sciences\mbox{,} University of Chinese Academy of Sciences\mbox{,} Beijing 100049\mbox{,} China\\
$^{10}$ Songshan Lake Materials Laboratory\mbox{,} Dongguan\mbox{,} Guangdong 523808\mbox{,} China\\
$^{*}$These authors contributed equally: Ge He and Shiyu Xie\\
\textrm{\Letter}Corresponding e-mails:
ge.he@bit.edu.cn;
jshen@bit.edu.cn; 
duzengyi@hfnl.cn;
dlfeng@hfnl.cn
}

\begin{abstract}
\section{Abstract}

The bilayer nickelate superconductor \LNO undergoes a density wave transition near 150 K that has attracted intensive scrutiny, yet its microscopic origin remains elusive. Here we report polarization-resolved electronic Raman scattering measurements on high-quality single crystals of \LNO. Below 150\,K, we observe a pronounced, symmetry-dependent redistribution of spectral weight in \Blg and \BZg channels, consistent with the formation of spin-density-wave (SDW) gaps. Quantitative analysis reveals momentum-selective SDW gap amplitudes, with intermediate-to-strong coupling near X/Y points of the Brillouin zone and weaker coupling along the diagonal direction, indicating an unconventional SDW driven by anisotropic electronic correlations. Our results establish the electronic character of the SDW in \LNO, and provide a microscopic foundation for understanding the emergence of high-temperature superconductivity under pressure in nickelates.

\end{abstract}
\maketitle

\section{Introduction}

\begin{figure*}[ht]
   \includegraphics[width=0.9\textwidth]{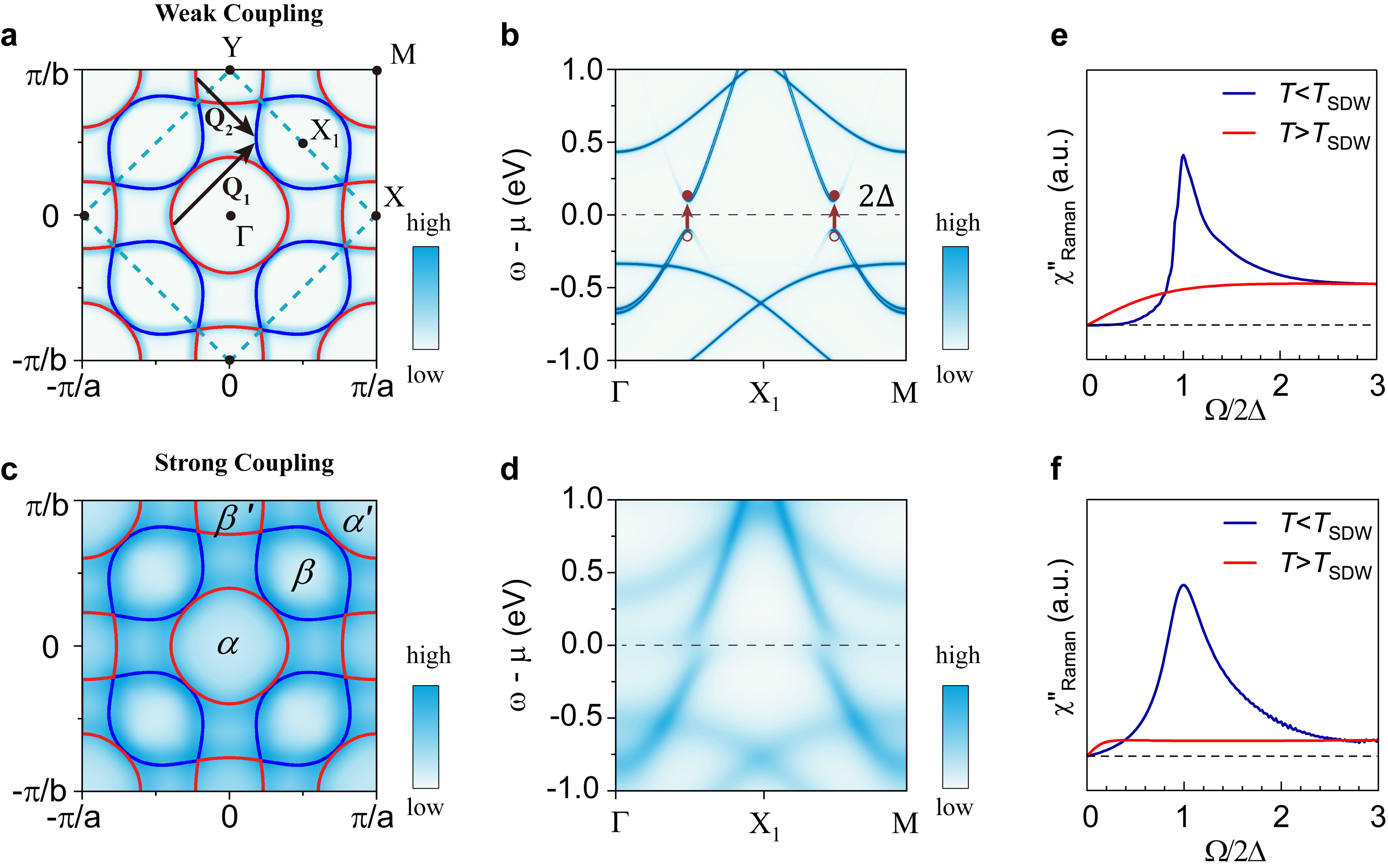}
   \caption{\textbf{Comparison of the electronic structure and Raman spectroscopic characteristics in weak and strong coupling regimes in \LNO.} \textbf{a}, Fermi surface in the weak coupling regime, calculated using an 8-band tight-binding model (see Supplementary Materials F for details). The solid lines represent the Brillouin zone (BZ) of the ideal unit cell (without considering tilted Ni-O octahedra), while the dashed lines indicate the BZ of the real unit cell (with tilted Ni-O octahedra). The blue and red curves denote hole ($\beta$) and electron ($\alpha$, $\beta'$, $\alpha'$) pockets arising from band 3 and band 4, respectively. The black arrows indicate the wavevectors \textbf{Q}$_1$ connecting $\alpha$ and $\beta$ pockets and \textbf{Q}$_2$ connecting $\beta$ and $\beta'$ pockets, respectively. \textbf{b}, Spectral weight $A(k,\omega)$ calculated within a mean-field approximation considering a density wave gap $\Delta$ induced by Fermi surface nesting with a specific wave vector. \textbf{c}, Fermi patches in the strong coupling regime, where strong interactions lead to a broadened distribution of electronic states near the Fermi level across the BZ.  \textbf{d}, Spectral weight $A(k,\omega)$ in the strong coupling regime, where a broad continuum appears due to incoherent particle-hole mixing, allowing excitations both below and above the Fermi level, unlike the sharp features in a normal band picture. \textbf{e}, Typical Raman spectral features of an SDW system in the weak coupling regime, exhibiting well-defined coherence peaks. \textbf{f}, Raman spectral response of an SDW system in the strong coupling regime, characterized by a broad redistribution of spectral weight instead of sharp features. The spectra below \TSDW are calculated using simulated spectral functions at the mean-field level or including correlation effects, while those above \TSDW are simulated with the memory-function approach using reasonable parameters. Details of the simulations are provided in Supplementary Materials E and G.}

   \label{fig:motivation}
\end{figure*}

\begin{figure*}[ht]
   \includegraphics[width=1\textwidth]{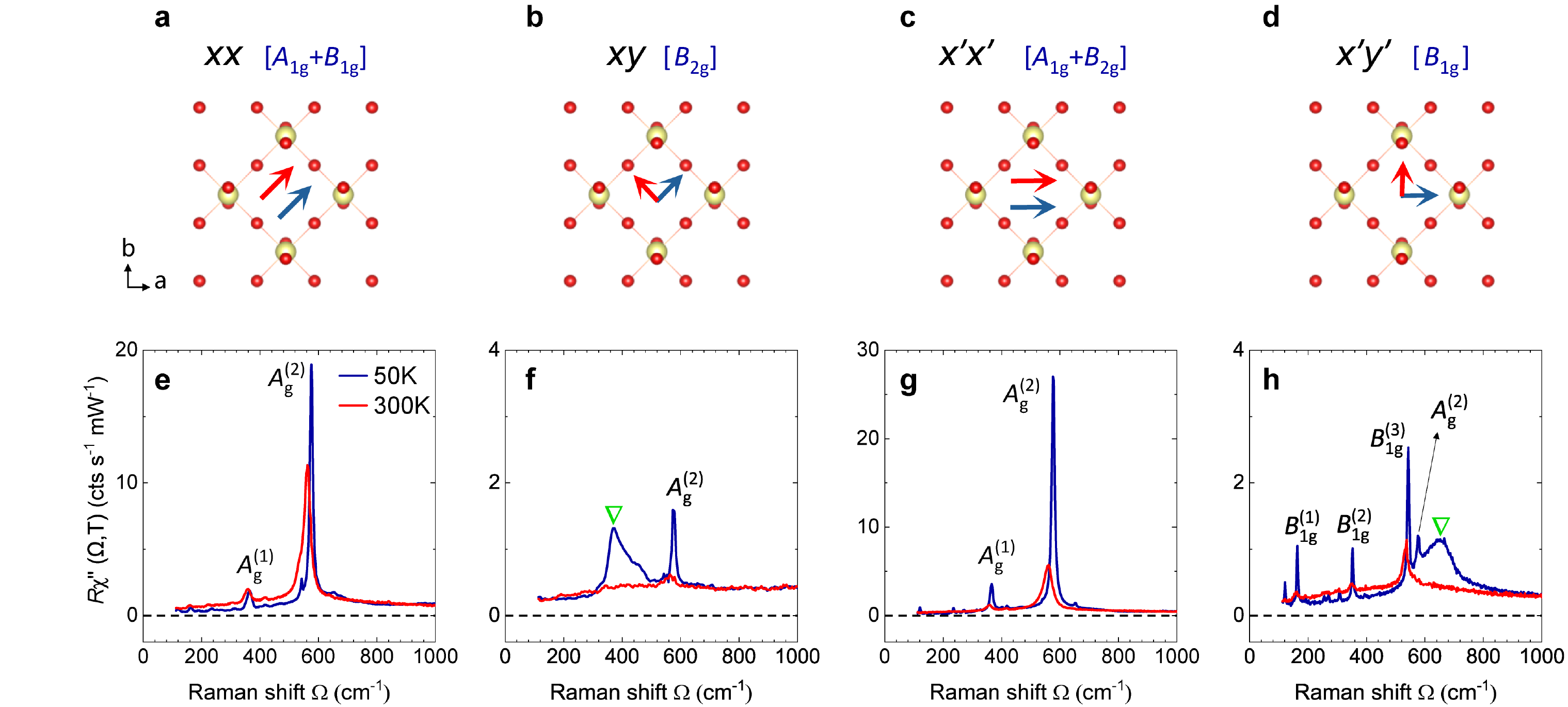}
   \caption{\textbf{Polarization configurations and corresponding Raman spectra.} \textbf{a}–\textbf{d}, Schematic definitions of the polarization configurations, where red and yellow spheres represent oxygen and nickel atoms, respectively. The $x$- and $y$-axes are aligned along the Ni–O–Ni bond directions. The $x'$ and $y'$ polarizations are rotated by 45$^\circ$ clockwise from the $x$ and $y$ axes, respectively. The symmetry channels shown in bule are defined within $D_{4h}$ point group. \textbf{e}–\textbf{h}, Raman spectra measured at 50\,K and 300\,K for the corresponding configurations. Raman-active phonon modes are labeled as indicated. Green triangles denote the presence of additional electronic Raman responses in the $xy$ and $x'y'$ channels.}

   \label{fig:rawdata}
\end{figure*}

High-temperature superconductivity is often observed in proximity to an antiferromagnetic (AFM) order, as seen in cuprates \cite{Keimer:2015}, iron pnictides and chalcogenides \cite{Dai:2015}, and more recently, the Ruddlesden-Popper (RP) phase of nickelates \cite{Wang:2024}. This recurring association suggests that magnetism may serve as a common thread in the quest for microscopic origin of high-temperature superconductivity \cite{Armitage:2010, Kivelson:2003, Scalapino:2012}. 

The bilayer RP phase of \LNO has been found to exhibit high temperature superconductivity above 77 K under high pressure \cite{Sun:2023, Zhang:2024, WangbulkSC:2024, Zhu:2024,Shimengzhu:2025,Zhangjunjie:2025}. At ambient pressure, \LNO exhibits density-wave (DW) like transitions near 150 K, which have been extensively studied by multiple experimental techniques \cite{Sreedhar:1994,Goodenough:19944,Taniguchi:1995,Kobayashi:1996,LING:2000517,Wuguoqing:2001,Liu2022,Zhangjunjie:2023,Liuzhe:2024,RIXS:2024,Shulei:2024,XIE:20243221,Hawthorn:2024,Yuxiaohui:2024,Kakoi:2024,Khasanov:2025,WutaoNMR:2025,Yanglexian:2025,LaPrNO:2025}. While optical measurements, including ultrafast and infrared spectroscopy \cite{Yuxiaohui:2024,Liuzhe:2024}, reported gap-opening signatures associated with DW transitions, angle resolved photoemission spectroscopy (ARPES) measurements on \LNO show no clear gap opening \cite{Zhouxingjiang:2024,ARPESYanglexian:2024,ZXSHen:2025,Damascelli:2025}. Resolving this experimental dichotomy between optical and ARPES observations may provide key insights into the microscopic nature of DW ordering. 

Spin density wave (SDW) and charge density wave (CDW) orders have been shown to be closely intertwined \cite{Liu2022,RuiZhou:2025,Khasanov:2025} with the SDW component exhibiting a larger amplitude, as evidenced by resonant inelastic X-ray scattering (RIXS) \cite{RIXS:2024}, neutron scattering \cite{XIE:20243221}, $\mu$SR \cite{Khasanov:2025} and nuclear magnetic resonance (NMR) \cite{WutaoNMR:2025} experiments. Efforts to elucidate the nature of SDW in \LNO fall into two regimes, depending on the underlying coupling strength. In the weak coupling regime \cite{Greenblatt:1996,Hujiangping:2024,Zhangyang:2024,Ilya:2024,Yangfan:2023,Ouyang:2024}, density-wave instabilities are driven by Fermi surface nesting \cite{Zhouxingjiang:2024,ARPESYanglexian:2024,ZXSHen:2025}. In this context, possible nesting wavevectors, denoted \textbf{Q}$_1$ and \textbf{Q}$_2$, are superimposed on the calculated Fermi surface as shown in Fig. \ref{fig:motivation} \textbf{a}. Consequently, SDW gaps open around the Fermi energy ${E_F}$ on these nested Fermi surfaces in conjunction with the density wave transition (Fig. \ref{fig:motivation} \textbf{b}). 

In the strong coupling regime, by contrast, localized spin orders are stabilized \cite{Yaodaoxin:2023,Zhangyang:2023,Hujiangping:2023,Wangqianghua:2023,Wucongjun:2024,Sugang:2024,Yangyifeng:2024,Liwei:2024} driven by strong correlation effect and Hund's coupling \cite{Philipp:2023,Yangyifeng:2024,Luzhongyi:2024,Gabriel:2025}. This scenario is strongly supported by RIXS measurements, which directly reveal an SDW-type magnetic excitation \cite{RIXS:2024}. Notably, the dispersive magnetic excitations soften to zero energy at the wavevector ($\pi$/2, $\pi$/2), indicating the formation of quasi-static spin order near $T_{\text{SDW}}$$\approx$150 K. As a consequence, the spectral function becomes markedly incoherent in the strong-coupling regime, with the Fermi surface appearing substantially broadened across the BZ (Fig. \ref{fig:motivation} \textbf{c}), distinct from the sharp, coherent Fermi surface depicted in Fig. \ref{fig:motivation} \textbf{a}. Simultaneously, spectral weight spreads broadly across the phase space (Fig. \ref{fig:motivation} \textbf{d}), accompanied by a transfer of spectral weight between particle-like and hole-like excitations across the Fermi level, and the emergence of partial gaps associated with the SDW transition.

Raman spectroscopy, which probes both particle–particle and particle–hole excitations at $\textbf{q} \rightarrow 0$, is a powerful tool for accessing electronic states across different regions of the Brillouin zone via polarization selection rules~\cite{Devereaux:2007}. Importantly, the Raman susceptibility is sensitive to the strength of electronic interactions. In the weak-coupling regime, the Raman response typically exhibits a sharp 2$\Delta$ singularity with an extended high-energy tail (Fig. \ref{fig:motivation} \textbf{e}). In contrast, strong-coupling behavior leads to a broad continuum-like response, often lacking a distinct singularity and manifesting as a broad peak, which may or may not be symmetric (Fig. \ref{fig:motivation} \textbf{f}, details of the simulations producing these curves are provided in the Supplementary Materials G). To resolve the nature of SDW order in \LNO, we performed polarization-resolved electronic Raman measurements to distinguish between weak- and strong-coupling regimes. Below the transition temperature, $T < $ \TSDW, the Raman response of \BZg channel exhibits a sharp coherence peak with asymmetric lineshape. In contrast, the \Blg channel displays a broad, incoherent, and symmetric-like peak. The temperature dependence of both channels is closely correlated with the SDW transition at $T_{\text{SDW}} \approx150$~K. Guided by Raman selection rules, \Blg and \BZg spectra primarily probe electronic states near X/Y points and along the diagonal direction respectively. The extracted gap magnitudes are $\Delta_{B_{\rm{1g}}}$ = 37.5-40.4 meV, and $\Delta_{B_{\rm{2g}}}$ = 23.0 meV, corresponding to gap ratios of 2$\Delta_{B_{\rm{1g}}}/k_B T_{\text{SDW}} \approx$ 5.5-5.9 and 2$\Delta_{B_{\rm{2g}}}/k_B T_{\text{SDW}} \approx$ 3.4. These values indicate weak and medium-to-strong coupling SDW mechanisms across various regions of the Brillouin Zone (BZ). Thus, these results establish Raman spectroscopy as a sensitive probe of SDW electronic nature in \LNO and provide critical insight into the microscopic origin of density-wave formation in this single crystal.

\section{Results}


\LNO single crystals belong to the $D_{\rm{2h}}$ point group \cite{Sun:2023}. Factor group analysis predicts ten \Ag and twelve \Blg phonons to be Raman-active for light polarizations within the $ab$-plane (see Supplementary Materials B for details). We identified two \Ag and three \Blg prominent phonon modes, labeled $A_{\rm g}^{(1)}$, $A_{\rm g}^{(2)}$, and $B_{\rm {1g}}^{(1)}$–$B_{\rm {1g}}^{(3)}$, respectively (Fig.~\ref{fig:rawdata}). Details regarding the identification of the weaker modes and the temperature dependence of their frequencies and linewidths are provided in Supplementary Materials B. A similar Raman spectrum of \LNO, without polarization and temperature dependence, has been previously reported in Refs.~\cite{Yuxiaohui:2024,Yanglexian:2025}. In addition to phonon modes, we observed broad peaks marked by green triangles, located at approximately 370\,\wn and 650\,\wn in the $xy$ and $x'y'$ channels, respectively (see Fig.~\ref{fig:rawdata} \textbf{f} and \textbf{h}). These features originate from electronic Raman scattering.

\begin{figure}[t]
   \includegraphics[width=0.45\textwidth]{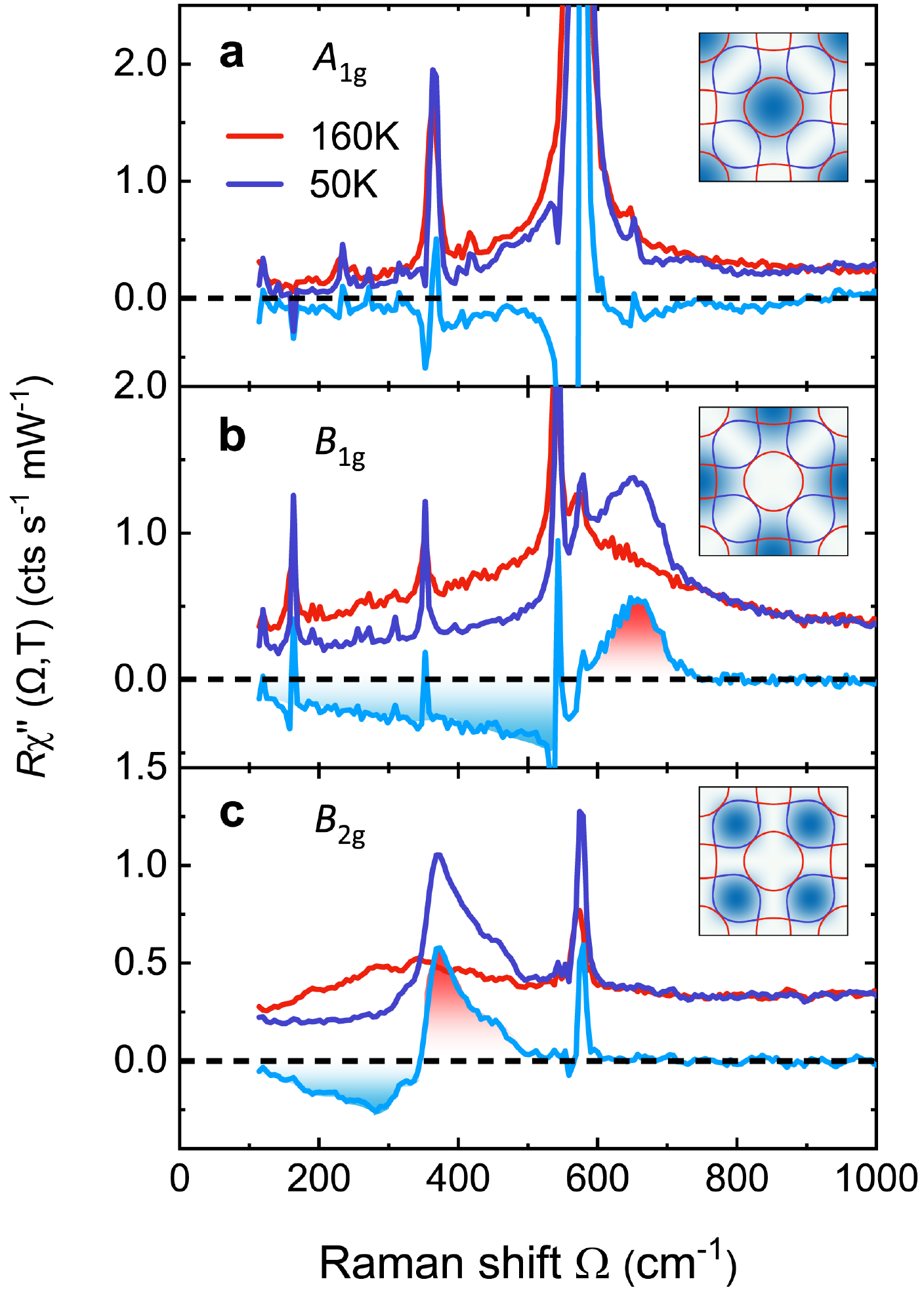}
  \caption{\textbf{Normal and SDW state Raman spectra of \LNO at temperatures as indicated.} The difference spectra between 50\,K and 160\,K are overlaid as light blue curves. Spectral weight redistribution is clearly observed in the \Blg and \BZg channels. The spectral weight loss is highlighted in blue, while the gain is indicated in red. Insets: Color maps of Raman vertices in the first BZ for the \Alg, \Blg, and \BZg symmetries, respectively. }

   \label{fig:continuum}
\end{figure}

To better understand the selection rules governing the electronic Raman response, we adopt a pseudo-tetragonal point group symmetry, $D_{\rm{4h}}$, which yields three relevant irreducible representations: \Alg, \Blg, and \BZg. The validity of the pseudo-$D_{\rm{4h}}$ approximation is examined in Supplementary Materials C. The corresponding Raman vertices, which are proportional to the crystal harmonics \cite{Devereaux:2007, Mijin:2021}, are illustrated in the insets of Fig.~\ref{fig:continuum}. These vertices highlight the momentum-space regions where particle-hole excitations are selectively probed: \Alg ($x^2+y^2$) projects electronic states near the BZ center and corners, \Blg ($x^2-y^2$) predominantly selects the excitations around the X/Y points, and \BZg ($xy$) emphasizes the electronic states along the diagonal direction.

Figure~\ref{fig:continuum} presents the electronic Raman responses in the \Alg, \Blg, and \BZg channels of \LNO over the range 100–1000\,\wn at 50\,K and 160\,K. The difference spectra [$\chi''$(50\,K) - $\chi''$(160\,K)] are shown as light blue curves. A clear redistribution of the the spectral weight is observed in both the \Blg and \BZg symmetries, whereas no significant redistribution is found in the \Alg spectra. Note that the \Alg spectra are extracted using a linear combination of parallel and cross configurations (see more details in Supplementary Materials D). The dips at 390\,\wn and 570\,\wn in the \Alg difference spectra arise from temperature-induced changes in phonon modes (see Fig.~\ref{fig:continuum}\textbf{a}). In the \Blg channel, a pronounced redistribution of spectral weight emerges characterized by a spectral weight loss below 600\,\wn and a corresponding gain between 600 and 720\,\wn (Fig.~\ref{fig:continuum}\textbf{b}). The line shape is nearly symmetric. In contrast, the \BZg channel displays an asymmetric peak with a tail at the high energy side near 370\,\wn at low temperature (Fig.~\ref{fig:continuum}\textbf{c}).

\begin{figure*}[ht]
   \includegraphics[width=1\textwidth]{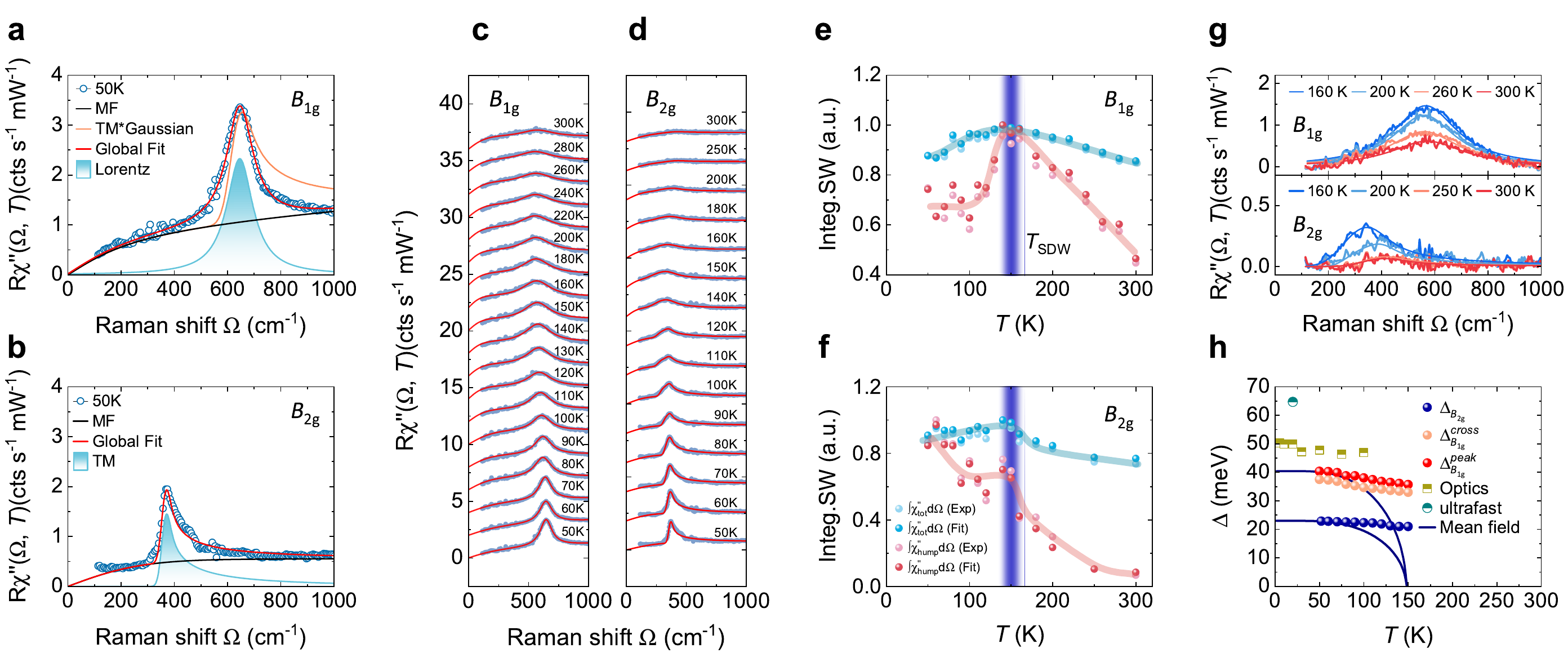}
   \caption{\textbf{Spectral weight and energy gap in \LNO.} 
   \textbf{a,b}, Fits of the electronic continuum at 50~K using phenomenological memory function (MF)–Tsuneto-Maki (TM) and MF–Lorentz models for the \Blg and \BZg spectra, respectively. A TM function convoluted with a Gaussian \cite{Burch:2007} is also used to fit the \Blg spectrum (yellow line). 
   \textbf{c,d}, Electronic Raman responses (gray points) and their corresponding fits (red curves) at various temperatures.
  \textbf{e,f}, Integrated spectral weight of the total spectrum and SDW component for both the experimental data and fitted results from 100 to 1000~\wn as a function of temperature in the \Blg\ and \BZg channels. 
   The SDW transition temperature is marked by the light-blue vertical bands. 
   \textbf{g}, Raman response in the \Blg (top panel) and \BZg (bottom panel) symmetry at 160, 200, 260, and 300~K. For clarity, the phonon peaks are subtracted from the spectra.
   \textbf{h}, Temperature dependence of the SDW energy gaps. The half-filled circles and squares are adapted from ultrafast spectroscopy~\cite{Yuxiaohui:2024} and optical conductivity~\cite{Liuzhe:2024}, respectively. Blue and red (yellow) points represent the energy gaps measured in the \BZg and \Blg symmetries, respectively. The deviation from the mean-field prediction is illustrated by the blue curves.}
   \label{fig:SW}
\end{figure*}


To further investigate the redistribution of spectral weight, we perform a quantitative analysis of the electronic continuum, as shown in Fig.~\ref{fig:SW}. Before the analysis, all phonon lines were fitted using Voigt functions and subsequently subtracted from the spectra. The background scattering is modeled using a memory function method \cite{Gotze:1972, Opel:2000, Sen:2020} 

\begin{equation}
\chi''_{\rm MF}(\Omega) = \frac{\hbar\Omega\Gamma(\Omega)}{[\hbar\Omega(1+\lambda(\Omega))]^2+[\Gamma(\Omega)]^2},
\end{equation}

\noindent where $\Gamma(\Omega)$ is the dynamic scattering rate of the charge carriers, $\lambda(\Omega)$ is the Kramers-Kronig (KK) transformation partner of $\Gamma(\Omega)$ (see Supplementary Materials E for details).

Given the distinct lineshapes of the electronic Raman responses in the \Blg and \BZg symmetries, we adopt two different models for fitting. For asymmetric lineshapes, we use the Tsuneto-Maki (TM) function~\cite{Tsuneto:1960}, which is typically applied in superconducting systems but also applicable to density-wave states~\cite{Lazarevic:2020}:

\begin{equation}
\chi''_{\rm TM}(\Omega) = \frac{\pi}{2}\frac{(2\Delta)^2}{\Omega\sqrt{\Omega^2-(2\Delta)^2}}, \quad \Omega > 2\Delta,
\end{equation}

\noindent where $\Delta$ denotes the density-wave gap.

For nearly symmetric lineshapes, the inelastic Raman response is empirically modeled by a Lorentzian function:

\begin{equation}
\chi''_{\rm Lorentz}(\Omega) = \frac{2A}{\pi}\frac{\Gamma}{4(\Omega-\Omega_0)^2+\Gamma^2},
\end{equation}

\noindent where $A$ is the resonance amplitude, $\Omega_0$ is the resonance frequency, and $\Gamma$ is the linewidth. Note that the Lorentzian function describes an isolated oscillator and therefore does not incorporate coherence effects characteristic of superconductivity~\cite{He:2020} or SDW transitions~\cite{Gallais:2011}. It is employed here purely as a phenomenological tool to extract the peak position and integrated intensity.

We fit the \Blg and \BZg spectra using $\chi'' =  \chi''_{\rm MF}$+ $\chi''_{\rm Lorentz}$ and $\chi'' =  \chi''_{\rm MF}$+ $\chi''_{\rm TM}$, respectively. The contributions from each fitting component at 50~K are shown in Fig.~\ref{fig:SW}\textbf{a} and \textbf{b}. The temperature-dependent electronic Raman responses together with their corresponding fits are presented in Fig.~\ref{fig:SW}\textbf{c} and \textbf{d}. The fits reproduce the experimental data well. More clearly separated fits at various temperatures can be found in Figs.~S5 and S6 of the Supplementary Materials E.

To further disclose the nature of the SDW phase transition, we investigated the corresponding integrated spectral weights for both the entire electronic Raman response and the isolated SDW contribution, as shown in Fig.~\ref{fig:SW}\textbf{e} and \textbf{f}.
A transition temperature around 150\,K is clearly identified in both symmetries. 
For the integration over the whole electronic response, the spectral weight exhibits a cusp-like temperature dependence in both \Blg and \BZg symmetries. However, for the integration of the pure SDW response, the behaviors differ significantly.
In the \Blg symmetry (Fig.~\ref{fig:SW} \textbf{e}), the spectral weight first increases and then decreases with rising temperature. In contrast, in the \BZg symmetry (Fig.~\ref{fig:SW}\textbf{f}), the spectral weight decreases with a kink at \TSDW. At 50\,K, a total spectral weight loss of up to 10\% of the maximum intensity is observed in both the \Blg and the \BZg symmetries. Unlike optical conductivity, Raman scattering does not obey a sum rule, and therefore the spectral weight loss and gain are not required to balance each other~\cite{Kosztin:1991}. 
The observed transition is attributed to SDW ordering, which will be discussed in detail in the Discussion section.

Figure \ref{fig:SW} \textbf{g} presents the spectra above 150\,K in the \Blg and \BZg symmetry. A residual intensity peaked at approximately 570\,\wn and 380\,\wn is observed, respectively. Unlike the spectral weight redistribution below 150\,K, where both gain and loss of the spectral weight are evident, the high-temperature behavior is characterized solely by an intensity gain. Additionally, the peak intensity gradually decreases as the temperature increases. This behavior can be well reproduced across various samples, indicating an intrinsic property of \LNO (see Supplementary Materials H for details).

The temperature-dependent SDW gap is plotted in Fig.~\ref{fig:SW}~\textbf{h}. The gap size measured from the \BZg spectra is extracted directly from the fits, with a maximum value of $\mathit{\Delta}_{B_{\rm{2g}}}$ $\approx 23$\,meV. In contrast, the maximum gap measured from the \Blg spectra, estimated from the crossing point (and peak) between spectra measured at 50\,K and at 160\,K, is $\Delta^{cross}_{B_{\rm{1g}}}$ $ \approx 37.5$\,meV ($\Delta^{peak}_{B_{\rm{1g}}}$ $\approx 40.4$ \,meV). These correspond to ratios of 2$\Delta_{B_{\rm{1g}}}/k_B T_{\text{SDW}} \approx$ 5.5-5.9 and 2$\Delta_{B_{\rm{2g}}}/k_B T_{\text{SDW}} \approx$ 3.4, respectively. The lineshape of the Raman response provides key insights into the underlying SDW gap structure. The SDW gap in \BZg channel is isotropic, whereas the gap observed in \Blg channel exhibits slight anisotropy. The gap anisotropy is manifested as a finite low-energy slope in the \Blg response, whereas the \BZg symmetry exhibits an almost flat behavior. Notably, the temperature dependence of both gaps is significantly weaker than predicted by mean-field theory (see blue curves in Fig.~\ref{fig:SW}~\textbf{h}). For comparison, the SDW gaps obtained from Infrared~\cite{Liuzhe:2024} and ultrafast optical spectroscopy~\cite{Yuxiaohui:2024} are also plotted in Fig.~\ref{fig:SW}~\textbf{h}. These values are higher than those observed in our Raman measurements, which may be attributed to differences in probe sensitivity and/or variations in oxygen content across samples~\cite{Dong:2024}. To sum up, our observation reveals two distinct SDW gaps opened on different regions of the BZ characterized by different gap magnitudes, coupling strengths, and even gap structures. These findings underscore the anisotropic SDW nature in \LNO.

\section{Discussion}

We performed systematic Raman spectroscopy on \LNO at ambient pressure. First, no anomalies were observed in the phonon modes across the transition, arguing against a dominant CDW instability (see Supplementary Materials B), consistent with previous studies reporting a magnetic transition near 150 K~\cite{RIXS:2024, Shulei:2024, WutaoNMR:2025}. Second, we detected a redistribution of spectral weight in the electronic Raman continuum. In general, spectral weight redistribution may originate from gap opening~\cite{Eiter:2013, He:2024,Gallais:2011,Devereaux:1994}, strong electronic correlations~\cite{Freericks:2001, Nyhus:1995, Chen:1997, Sen:2020, Ye:2022}, Kondo lattice~\cite{Ye:2022} or band‐structure reconstructions~\cite{LeBlanc:2010}. In contrast to the broad spectral changes without a sharp threshold, typically arising from correlation or band effects, the SDW gap opening produces a distinct suppression of low-energy excitations and a sharp enhancement of spectral weight near the gap edge ($2\mathit{\Delta}$). Thus, we associate the spectral weight redistribution with the opening of an SDW gap, a characteristic feature also observed in the iron pnictide BaFe$_2$As$_2$~\cite{Gallais:2011}. While the presence of coupled spin–charge ordering cannot be entirely excluded, our findings, together with earlier reports, indicate that SDW formation is the primary electronic instability in the normal state of \LNO~\cite{RIXS:2024}.

The SDW in \LNO\ has been proposed to originate from Fermi surface nesting with a wave vector \textbf{Q}$_1$, connecting the $\alpha$ and $\beta$ pockets, as suggested by Wang \etal~\cite{Hujiangping:2024}. Alternatively, a nesting scenario involving a wave vector \textbf{Q}$_2$, connecting the $\beta'$ and $\beta$ pockets, has been supported by the observation of a 'translated' $\beta$ Fermi surface, consistent with scattering processes involving \textbf{Q}$_2$
\cite{Damascelli:2025}. The \textbf{Q}$_1$ scenario would imply a comparable gap opening on the $\alpha$ pocket, which should, in principle, be observable in the \Alg Raman spectra. However, this is not straightforward. Screening effects can suppress the Raman intensity of gap-related excitations, particularly in conventional metals~\cite{Devereaux:2007}. If the screening is negligible, as observed in \BKFA~\cite{Bohm:2014}, then the absence of gap signatures in the \Alg spectra would argue against the nesting vector \textbf{Q}$_1$.

In contrast, the observed gap features in the \Blg and \BZg spectra may support wavevector \textbf{Q}$_2$, connecting the $\beta'$ and $\beta$ pockets. The distinct spectral lineshape and different $2\mathit{\Delta}/k_B T_{\text{DW}}$ ratio, however, indicate an unconventional SDW microscopic mechanism in \LNO. Specifically, the gap opened close to X/Y points shows a medium-to-strong coupling strength, while along the diagonal direction, the gap exhibits weak coupling. Notably, such coupling strength cannot be accounted for by a simple weak-coupling picture with slight gap anisotropy along the diagonal direction, as evidenced by the failed fitting using the TM function convoluted with a Gaussian peak \cite{Burch:2007} showing in Fig.~\ref{fig:SW}\textbf{a}. Such strong coupling can lead to incoherency of the bands, and partial gap opening over a large momentum space in ARPES measurements, instead of clean leading-edge gap, as reported in 2H-Na$_x$TaS$_2$~\cite{Shen:2007} and NbSe$_2$~\cite{Shen:2008}. This may account for the absence of a well-defined gap in ARPES measurements \cite{Zhouxingjiang:2024,ARPESYanglexian:2024, ZXSHen:2025, Damascelli:2025}. However, the weak-coupling gap, which should in principle be observable by ARPES, has not yet been detected. This discrepancy may stem from the differing sensitivities of the probes. Raman spectroscopy is bulk-sensitive, whereas ARPES primarily probes the surface, potentially leading to divergent observations. Note that optical conductivity reports only a single gap near 50~meV \cite{Liuzhe:2024}. This difference may be attributed to the distinct nature of the optical conductivity and Raman responses. The former probes current–current correlations, whereas the latter measures effective density–density correlations, which can behave differently in correlated systems \cite{Devereaux:2007, Basov:2005}. In addition, Raman selection rules emphasize excitations in specific BZ regions, whereas optical conductivity averages over the entire zone, potentially masking anisotropic gap features.
The variation in coupling strengths across different pockets suggests the presence of anisotropic coherency of the quasiparticles in \LNO, a phenomenon widely reported in Fe-based superconductors \cite{Haule:2009, Liu:2015} and often attributed to strong Hund’s coupling effects involving multiple orbitals. The anisotropic SDW correlations are also reflected in the distinct behavior of the integrated spectral weight in the \Blg\ and \BZg\ symmetries (Fig.~\ref{fig:SW}\textbf{e} and \textbf{f}).

Moreover, residual peaks centered at approximately 570 \wn (71~meV) and 350 \wn (43~meV) are observed in the \Blg and \BZg spectra above $T_{\mathrm{SDW}}$, respectively. Their intensity gradually decreases with increasing temperature. In contrast, the peak positions exhibit distinctly different temperature evolutions above and below $T_{\mathrm{SDW}}$. Above $T_{\mathrm{SDW}}$, the peak position remains nearly constant in the \Blg\ channel and even shifts slightly toward higher energy in the \BZg\ channel. Such behavior is not characteristic of a conventional gap opening. Instead, it may reflect short-range magnetic correlations, Lorentzian-type spin fluctuations, or correlation-induced pseudogap behavior, which cannot be conclusively distinguished at present. Notably, RIXS and NMR measurements report that the existence of short-ranged spin correlations above $T_{\text{SDW}}$ and diminishes gradually with increasing temperature~\cite{RIXS:2024,WutaoNMR:2025}, although the onset temperature inferred from those probes appears slightly lower than that observed in our Raman measurements. Additionally, the deviation of the gap's temperature dependence from mean-field behavior further highlights the unconventional nature of the SDW gaps, possibly indicating the presence of spin fluctuations above $T_{\mathrm{SDW}}$ \cite{Gruner:1994}.

Recent $\mu$SR experiments have reported that the SDW order in \LNO is enhanced under increasing pressure~\cite{Khasanov:2025}. This finding highlights the importance of exploring the pressure dependence of the electronic Raman response. High-pressure Raman measurements would directly track the evolution of the SDW gaps, offering a valuable spectroscopic probe of the underlying electronic structure changes. Such investigations could provide critical insights into the interplay between magnetism and superconductivity in high-$T_c$ nickelate superconductors, potentially uncovering key mechanisms that drive the emergence of superconductivity in these complex materials.\\

\noindent\textbf{Methods}\\
\noindent\textbf{Samples:} Single crystals of La$_3$Ni$_2$O$_7$ were grown using a high-pressure optical floating zone furnace (model HKZ, SciDre GmbH). The precursor powder was synthesized by the standard solid-state method. The raw materials of La$_2$O$_3$ were dried at 1000 °C for 12 hours before use. Stoichiometric amounts of La$_2$O$_3$ and NiO powder were mixed and sintered at 1200°C for 1 day. Subsequently, the black powders were hydrostatically pressed into rods with 5 mm in diameter and 100 mm in length at 35,000 psi and then heated for 48 hours at 1200 °C. Single crystals La$_3$Ni$_2$O$_7$ were grown by the floating zone method in flowing O$_2$ atmosphere at a pressure of 15 bar with a flow rate of 0.2 L/min. During growth, seed rod were translated at a rate of 4 mm/h using a 5 kW Xenon arc lamp as a heating source. The quality of the samples was assessed by Laue diffraction, X-ray diffraction, and magnetization measurements, as detailed in Supplementary Material A. \\

\noindent\textbf{Light scattering:} The inelastic light scattering experiments were performed in a confocal geometry. The samples were mounted on the cold finger of a commercial Stirling fridge (ColdStation-50, MultiFields Tech.), allowing for temperature variation from 50\,K to 350\,K. A solid-state laser emitting at 532\,nm was used. In our experiments, the laser power was set at $P = 3.0$\,mW, resulting in a heating rate of $\sim$1\,K/mW. We present the Raman susceptibilities $R\chi^{\prime\prime}(\Omega,T) = \pi\{1+n(\Omega,T)\}^{-1}S(q \approx 0, \Omega)$ where $R$ is an experimental
constant, $\chi^{\prime\prime}$ is the imaginary part of Raman response function, $S(q \approx 0, \Omega)$ is the dynamical structure factor that is
proportional to the rate of scattered photons, and $n(\Omega,T)$ is
the Bose-Einstein distribution function \cite{Devereaux:2007}. To achieve high energy resolution for the phonon lines, we used a grating with 1800\,g/mm and a focal length of 800\,mm, resulting in an energy resolution of 1.66\,\wn. For measurements of the electronic continuum, we used a grating with 600\,g/mm, achieving an energy resolution of 5.90\,\wn to enhance the Raman intensity. \\

\noindent\textbf{Data Availability}\\
All relevant data that support the findings of this study are presented in the manuscript and supplementary information file. All data are available upon request from the corresponding authors.

\bibliography{refs}

\noindent\textbf{Acknowledgments}\\
We thank Rudi Hackl, Thomas P. Devereaux, George Sawatzky, Gheorghe Lucian Pascut,  Kejin Zhou, Xianxin Wu, Yaomin Dai and Tao Xiang for fruitful discussions.

\noindent\textbf{Funding}\\
This work was supported by the National Key Basic Research Program of China
(Grant No. 2024YFF0727103 (G.H.)), the National Science Foundation of China (Grant Nos. 12474473 (G.H.), U23A6015 (K.J.), 12104490 (Y.Z.), 12375331 (Z.Q.), 12425404 (M.W.), 52201036 (J.L.)), the New Cornerstone Science Foundation (Grant No. NCI202211 (D.F. and Z.D.)), the Innovation Program for Quantum Science and Technology (Grant No. 2021ZD0302803 (D.F. and Z.D.)), the Youth Beijing Scholars program (Grant No. 93 (L.Q.)), the Project of Construction and Support for high-level Innovative Teams of Beijing Municipal Institutions (Grant No. BPHR20220124 (L.Q.)), the CAS Project for Young Scientists in Basic Research (Grant No. 2022YSBR-048 (K.J.)), the Guangdong Basic and Applied Basic Research Foundation (Grant No. 2024B1515020040 (M.W.)), the Guangdong Provincial Key Laboratory of Magnetoelectric Physics and Devices (Grant No. 2022B1212010008 (M.W.)), the Research Center for Magnetoelectric Physics of Guangdong Province (Grant No. 2024B0303390001 (M.W.)), and the Synergetic Extreme Condition User Facility (SECUF, https://cstr.cn/31123.02.SECUF (G.H.)). J.Shen., S.X., H.Z., M.H., J.Shu., D.H., X.Z., L.Q., H.L., C.H., X.D., D.W., W.H., J.Y. and Y.Y. declare no relevant funding. \\

\noindent\textbf{Author contributions}\\
G.H., Z.D., J.Shen and D.L.F. conceived the project. G.H., S.Y.X., H.T.Z. and L.X.Q. performed the Raman measurements. C.S.H., H.J.L. and W.H. contributed to the experimental assistance. G.H., Z.D., J.Shu, Y.M.Z., X.X.Z., L.Q., X.J.D., D.J.W., J.L., Y.J.Y., Z.M.Q., J.Y. and K.J. analyzed the Raman data. M.W., M.W.H. and D.Y.H. synthesized and characterized the samples. G.H., Z.D. and Y.M.Z. wrote the manuscript with comments from all the authors.\\

\noindent\textbf{Competing interests}\\
The authors declare no competing interests.

\end{document}